\documentclass[num-refs]{nbdt-article}

\usepackage{siunitx}

\usepackage{amsmath, amsfonts}

\papertype{Original Article}
\title{Use and usability: concepts of representation in philosophy, neuroscience, cognitive science, and computer science}

\author[1,2\authfn{1}]{Ben Baker}
\author[3,4\authfn{1}]{Richard D. Lange}
\author[5\authfn{1}]{Andrew Richmond}
\author[6,7]{Nikolaus Kriegeskorte}
\author[8]{Rosa Cao}
\author[7,9,10]{Xaq Pitkow}
\author[11]{Odelia Schwartz}

\contrib[\authfn{1}]{Equally contributing authors, listed alphabetically.}

\makeatletter
\renewcommand{\Affilfont}{\fontsize{6\p@}{8pt}\selectfont\raggedright}
\renewcommand\AB@affilnote[1]{\footnotesize \textsuperscript{#1}}
\makeatother

\affil[1]{Department of Philosophy, Colby College, Waterville, ME, 04901, USA}
\affil[2]{Davis Institute for Artificial Intelligence, Colby College, Waterville, ME, 04901, USA}
\affil[3]{Department of Computer Science, Rochester Institute of Technology, Rochester, NY, 14623, USA}
\affil[4]{Center for Visual Science, Rochester, NY, 14627, USA}
\affil[5]{Rotman Institute of Philosophy, Western University, London, Ontario, N6A 3K7, Canada}
\affil[6]{Zuckerman Mind Brain Behavior Institute, Departments of Psychology and Neuroscience, Affiliated member of Electrical Engineering, Columbia University, New York City, NY}
\affil[7]{NSF AI Institute for Artificial and Natural Intelligence}
\affil[8]{Department of Philosophy, Stanford University, Stanford, CA, 94305, USA}
\affil[9]{Neuroscience Institute, Carnegie Mellon University, Pittsburgh, PA, 15213}
\affil[10]{Department of Machine Learning, Carnegie Mellon University, Pittsburgh, PA, 15213}
\affil[11]{Department of Computer Science, University of Miami, Miami, FL, 33146, USA}

\corraddress{Richard D. Lange, Computer Science, Rochester Institute of Technology, Rochester, NY, 14623, USA}
\corremail{richard.lange@rit.edu}

\runningauthor{Baker, Lange, Richmond et al.}

\begin{document}

\maketitle

\begin{abstract}
\small
Representations play a central role in the study of both biological and artificial intelligence, as well as philosophy of mind. Across neuroscience, computer science, and philosophy, a recurring theme is that representations not only carry information but should be ``useful'' for or ``usable'' by an agent in some sense. Here, we review how the ``usefulness'' of representations has been conceptualized and how it figures into different conceptions of representation. We identify and explore four aspects of use and usability: representations generally carry \textit{information}; that information may or may not be \textit{useful} and it may or may not be encoded in a usable \textit{format}; and the representations may or may not be \textit{used downstream}. Building on these four aspects of information and use, we then organize existing perspectives on neural representations into three levels: Representations as Information (Level 1); Representations as Usable (Level 2); and Representations as Used (Level 3). Our account is meant to give readers an appreciation for the diversity of notions of ``neural representation,'' help them navigate the vast and multi-disciplinary literature on the topic, and help them clarify the appropriate notion of representation for their own investigations.

% Please include a maximum of seven keywords
\end{abstract}

\section{Introduction}

Consider a person deciding which bunch of bananas to buy at the store, a rat navigating a maze in search of food, or a self-driving car identifying road signs from its cameras. In each case, an autonomous agent senses and interacts with its environment in a goal-directed fashion. We often think of such agents as internally ``representing'' parts of the world around them and using these representations to make decisions and guide behavior. The ways the person’s brain, the rat’s brain, and the car’s software represent their respective environments seem plausibly related. What these agents represent and how may be quite different, but the idea that either a biological or artificial agent ``represents'' is a core concept employed in multiple fields.

The study of representations is central to several otherwise quite distinct fields: neuroscience, computer science, and philosophy of mind (sometimes we will use ``cognitive science'' as an umbrella term). What exactly is a representation? This is a contentious question, and answers may vary depending on the system under study (e.g. human, rat, car), on who is studying it (e.g. scientist, engineer, or philosopher), and on what their goals are (e.g. predicting, creating, explaining) \citep{fallonWhatAreWe2023, bakerThreeAspectsRepresentation2022, favelaInvestigatingConceptRepresentation2023, richmond_commentary_2023, caoPuttingRepresentationsUse2022, eganDeflatingMentalRepresentation2025}. With such diverse contexts and goals, there may be no single universally agreeable concept of representation, even to the different authors of this paper. Nonetheless, comparing and contrasting how representations are conceived of and employed in these different fields is important. This can render previously tacit assumptions explicit, clarify agreements and disagreements, build bridges between fields, and open new lines of inquiry. We share the hope that the framework emerging from this collaboration provides a useful vocabulary, especially for clarifying meta-scientific points about how the term ``representation'' is used in different fields. 

Our goal in this work is to distill ideas from each field that we believe will be of interest to audiences across fields. We organize our inquiry around the principles of ``use and usability'' --- a common theme that has emerged in all fields interested in representations. For instance, some philosophers distinguish \textit{bona fide} representations from mere information-carrying states based on whether that state is ``used'' in behavior. Computer scientists might use the term representation more freely but are interested in constructing ``useful'' representations that enable a system to perform well on a task or set of tasks. This paper examines how such notions of ``use'' and ``usefulness'' are defined and how they are related.

\begin{figure}[bt]
\centering
\includegraphics[width=0.99\textwidth]{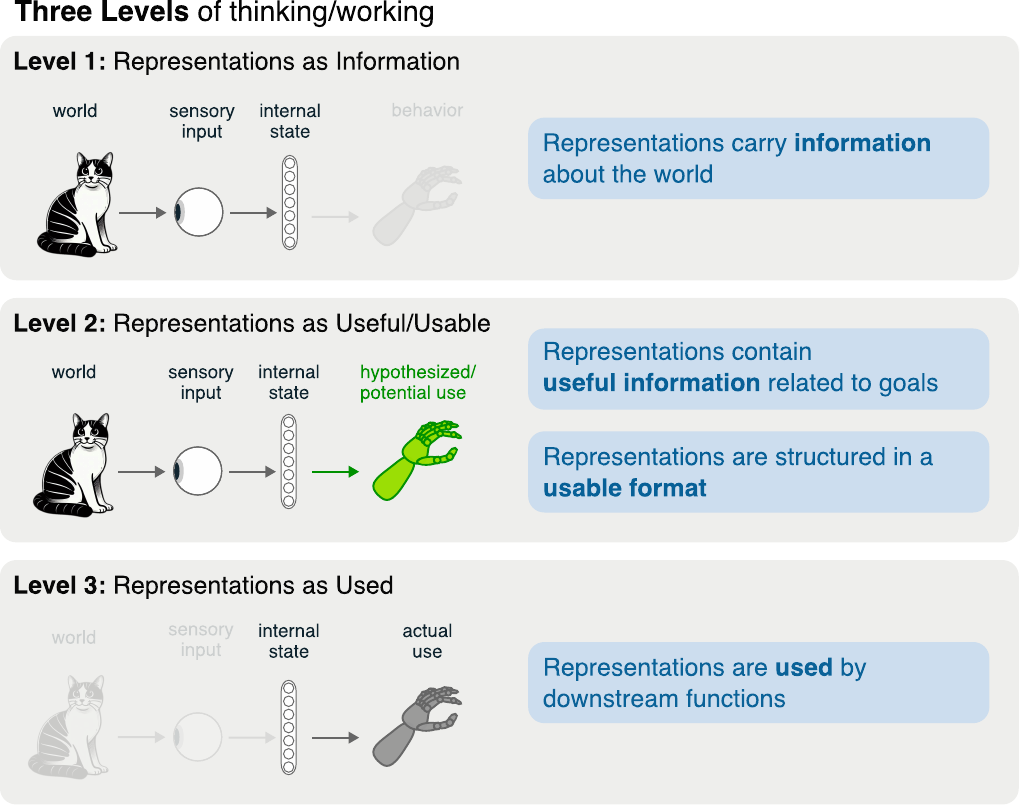}
\caption{High-level overview of the different \textbf{aspects} of ``use and usability'' (blue boxes, right) and how they are scaffolded in three \textbf{levels} of thinking/working with ``representations'' across fields. At each level, different emphasis is placed on the relationship between the world (e.g. a cat), how it affects internal states, and how these states might affect downstream behavior (e.g. petting the cat).}
\end{figure}

\section{Aspects of use and usability in defining representations}

\subsection{Preliminaries}

What exactly defines a representation of the kinds suggested by the examples above? This is not a question that we aim to answer univocally; in fact, any attempt to find a single definition would undermine our goal of comparing and contrasting how the concept of representation is used in different fields and different projects. Instead, we will characterize representations in terms of a few basic qualities that may or may not be shared across fields. As a starting point, we take representations to be some configuration or state internal to an agent (biological or artificial) that has content or meaning, or refers to or stands in for a part of the environment or another internal state (or to a more abstract or relational entity, or even one that does not exist, like an imagined unicorn). This pattern or state is often, but not necessarily, thought of as a vector (e.g. of neural activity across a population of neurons). We will use the term ``internal state'' throughout as a neutral term to refer to measured states of the cognitive system that may or may not qualify as representations. Typical examples of an internal state include voxel activity from fMRI or the activation of hidden units in a neural network, but the boundary of the cognitive system demarcating ``internal'' states could be extended to include the body or even parts of environment \citep{clarkExtendedMind1998,varelaEmbodiedMindCognitive1993}. To summarize, we are primarily interested in the notion of representation as it pertains to measurable ``internal'' states in the kinds of intelligent neural systems studied across neuroscience, computer science, and philosophy of mind.

\subsection{Four high-level aspects of use and usability}

Across neuroscience, computer science, and philosophy, a recurring theme is that representations not only carry information but should be useful for or usable by an agent in some sense. Four high-level aspects of use and usability emerged from our interdisciplinary discussions: representations are internal states that (1) \textbf{carry information}, where (2) that information is \textbf{useful} and (3) encoded in a \textbf{usable format}, and the states may be (4) \textbf{actually used} (have a causal role) in the system. Let us begin with an introduction to these high-level aspects to motivate the later discussions. 

First, and most straightforwardly, internal states carry information when they correlate with or statistically depend on some features of the world.

Second, internal states are sometimes characterized in terms of whether or not they are \textbf{useful}, i.e, they contain information relevant to the system’s goals. Consider the example of a rat navigating a maze in search of food. The details of the texture of the maze walls generally do not make a difference to the rat, so we would say that --- at least in the context of the task of solving the maze --- internal states that carry detailed information about wall texture are not useful. However, those same details may become useful in the context of a different task, like finding a good place to climb out of the maze. So whether some internal state carries useful information depends on the agent’s current or long-term tasks. Consequently the extent to which an agent can change goals or take up new goals complicates hypotheses about representation. These considerations turn out to be important across fields.

Third, internal states are sometimes characterized in terms of whether they occur in a \textbf{usable format}, which concerns the ability of hypothetical or actual downstream systems to process them. Regardless of the information content --- whether it concerns the location of rewarding cheese or wall texture --- there are different formats in which the information might be encoded in neural activity or other physical substrates, and these may prove more or less usable downstream. For instance, information in a pattern of neural activity that is linearly decodable is generally thought to be more usable than information that is formatted in a way that is only accessible by a complex nonlinear decoder \citep{dicarloUntanglingInvariantObject2007,kriegeskorteInterpretingEncodingDecoding2019,ivanovaLinearRegressionMapping2022}.

Importantly, useful information and usable format are distinct concepts, and an internal state may have one but not the other. A classic example of useful information in an unusable format is object identity as encoded by the immediate output of the retina --- while information about object identity is present in the retinal signals and relevant to many tasks or goals, further processing is required to format that information in a way that makes it ``explicit,'' i.e. (linearly) decodable \citep{kriegeskorteInterpretingEncodingDecoding2019}. Conversely, an internal state may be in a usable format but not be useful, such as linearly decodable but task-irrelevant details about wall texture. As we will elaborate further below, whether each of these cases qualifies as a ``representation'' may depend on who you ask, or why you are asking.

Fourth, internal states are sometimes characterized in terms of whether or not they are \textbf{actually used} by the system, which concerns their causal effect on downstream processes. An ultimate causal effect on behavior could be either immediate in the situation or delayed through internal processing and memory storage. In such cases, the relevant causal effect may be the effect on memory systems. This fourth aspect is important because one can imagine some internal state that contains  highly goal-relevant information in an easily decodable format, but that goes unused in the system in which it occurs. In other words, some internal state can be both in principle useful and usable but not play a causal role determining behavior. Whether such a state qualifies as a ``representation'' may again depend on who you ask.

These four high-level aspects of representational use and usability --- \textbf{information, usefulness, usable format, and actual use} --- are important because they can be used to organize the different ways that representations are conceptualized, with different conceptions putting different emphasis on each aspect and, most importantly, serving different scientific purposes. Below, we will rely on these aspects to characterize approaches to representation across fields and research programs.

\section{Organizing levels of thinking about representations}

How are the concepts of usefulness, usable format, and actual use related, and how do they factor into ideas about representation across disciplines? In this section, we will introduce and explore three ``levels'' of thinking about representations that emerged from our cross-disciplinary discussions. Building on the high-level aspects of information, usefulness, usable format, and actual use introduced above, our aim here is to compare different ways the term ``representation'' is used across research programs. Starting with the broadest conception of representation at Level 1, we will show how the notion of representation — and its role in scientific work — changes as we take into account more aspects of use and usability.

We emphasize that the later levels are not necessarily more correct than the earlier levels, nor are later levels necessarily nested within earlier ones. Rather, later levels add and subtract certain considerations about what is relevant to representation (see Figure 1, and the subsections below about which questions are enabled or occluded by each Level). Thinking and operating at Level 1 enables certain discussions and analyses that are not possible at later levels, and vice versa. Rather, the different levels can be viewed as supporting different research projects.

\subsection{Level 1: Representations as Information}

The first and least restrictive concept of representations considers them to be any internal state that carries information about something, regardless of how useful or usable the information is, and regardless of whether it actually influences downstream behaviors. At Level 1, any (non-constant) function of some data counts as a representation of that data. Activity in the hidden layers of a neural network represents the input to the network. Any neural activity that correlates with (or has some linear or nonlinear statistical dependence on) a state of the world represents that state. One’s heart rate represents one’s level of excitement or anxiety.
According to this view, the principal requirement for some internal state $\mathbf{r}$ in a biological or artificial agent to qualify as a representation of some $\mathbf{x}$ in the world is that $\mathbf{r}$ carries information about $\mathbf{x}$, or, equivalently, that $\mathbf{r}$ be statistically dependent on $\mathbf{x}$ (and that dependence must not be spurious). 

\subsubsection{What does Level 1 let us do? What questions does it answer?}

First, Level 1 draws a boundary around states with the feature that typically matters most about representations. If internal state $\mathbf{r}$ carries no information at all about $\mathbf{x}$, there would be little point in thinking of $\mathbf{r}$ as a ``representation of'' $\mathbf{x}$. (Setting aside for now  misrepresentation and hallucination, which we will revisit in Levels 2 and 3 below). 

Second, Level 1 enables the computational explanations common in computer science. For instance, the activity in some hidden layer of a feedforward neural network is commonly referred to as a ``representation of'' the input to the network. Level 1 then allows us to study mathematical properties of representations without getting bogged down by questions about whether they \textit{really are} representations in some more technical sense. For example, while representations at Level 1 may be defined as any function of the data, some representations are nonetheless preferable to others, and it is possible and valuable to study properties like the \emph{disentanglement} of representations without arguing about whether that property is a prerequisite for using representational terminology \citep{bengioRepresentationLearningReview2013,eastwoodFrameworkQuantitativeEvaluation2018,higginsDefinitionDisentangledRepresentations2018}. As with entanglement, the information bottleneck principle holds that the best representations for a machine learning task are those that contain as much task-relevant information as the input (a property known as \emph{sufficiency}) but that do not contain any other information (a property known as \emph{minimality}) \citep{tishbyDeepLearningInformation2015,achilleEmergenceInvarianceDisentanglement2018}. At Level 1, some internal state $\mathbf{r}$ may fall short of this ideal of being a minimal and sufficient representation of $\mathbf{x}$ for some task, but it would be a representation nonetheless. So, while aspects of use and usability are relevant here, they do not play a role in constraining the concept of representation, as they will in Levels 2 and 3.

Third, Level 1 connects the concept of representation to sensory data. This may be required to make ``neural representations'' empirically tractable, and provides a way into philosophical and psychological discourse about the ``grounding'' of representations. If $\mathbf{r}$ is a function of some quantity in the world $\mathbf{x}$, then we can look for an empirical signature in the dependence of $\mathbf{r}$ on $\mathbf{x}$ \citep{pohlClarifyingConceptualDimensions2026}. When a researcher’s data set $\lbrace{(\mathbf{x}_i, \mathbf{r}_i)}\rbrace$ comes from interactions of an agent with the world, an observed statistical dependence of internal state $\mathbf{r}$ on $\mathbf{x}$ provides an initial foothold for the idea of grounding representations in reality. That being said, the function $\mathbf{x} 
\rightarrow \mathbf{r}$ describes only one aspect of representation grounding, which may also depend on the structure or format of representations, i.e. whether they are purely amodal symbols, or rather reflect the structure of their referents, of perceptual representations, or of the embodied activity of the agent \citep{Barsalou2010,haimoviciModalAmodalDistinction2018,weberGroundingActionRepresentations2012}.

Given these positive qualities, it should come as no surprise that the Level-1 conception of representation is fairly common in the machine learning community. Outside of machine learning, a neuroscientist might demonstrate that neural responses correlate with a stimulus property, and then interpret the neural code as a representation. A study of this type can be understood as operating at Level 1 empirically, although the representational interpretation is often taken to imply that downstream regions make use of the information the neurons encode (alluding to a later level). Researchers in both neuroscience and computer science may also be interested in the question of which task(s) a representation might serve, or whether it is learnable by a certain algorithm, and if so, how fast. Importantly, these questions begin to touch on topics related to use and usability, but at Level 1 they are not seen as constraints on what is considered a ``representation.''

\subsubsection{What does Level 1 miss? What questions does it fail to answer?}

First, Level 1 does not entail a principle for identifying \emph{misrepresentations}. For example, consider the visual experience of two people, one who looks at the world through prismatic glasses that flip the image entering their eyes and one who looks at the world normally. The person wearing the prisms may (at least initially) make all sorts of errors, like reaching for an object to the left when it is actually to the right. It is intuitive to say that this person made these errors because their visual system misrepresented the locations of things. However, the statistical dependence of internal state on visual stimulus would contain the same degree of information for both people. 
Thus, Level 1 provides no basis for attributing errors like this to misrepresentation (see also Figure \ref{fig:faces-misrepresentation} below).

Second, Level 1 does not in general allow for a clear answer to the question of what the content of a particular representation is. At Level 1, an internal state $\mathbf{r}$ could be said to represent a wide variety of different things $\mathbf{x}$, as long as there is a reliable correlation between $\mathbf{r}$ and $\mathbf{x}$. 
To use an example from neuroscience, simple cells in primary visual cortex are classically thought of as edge detectors, responding to an oriented brighter patch next to a darker patch. Such a neuron could be said to be representing an edge in the image, or the boundary between two objects in the world, or a small element of some visual texture. At Level 1, it plausibly represents all of these at once, so a more restrictive notion would be required if we wanted to connect the representation to specific tasks or capabilities of the system \citep[cf.][]{eganDeflatingMentalRepresentation2025}.

Third, Level 1 is typically applied in studying representations of things in the immediate environment of the agent, since Level-1 representations must be a function of something, and that something is usually taken to be a state of the world. But not all notions of ``representation'' invoked by cognitive scientists and philosophers involve contents that refer to an actual state of the world. E.g., imagination and hallucination are often understood as representing future or imaginary scenarios. It is not clear what sensory data these representations would be a function of, or what the function would be. Cases such as this may be more fruitfully studied under a Level-2 or -3 concept of representation.

Finally, Level 1 may be too permissive for those who are interested in the special role that representational states play in complex systems like the brain. In cognitive science and philosophy, representations are typically understood as more than mere information-carriers \citep{ramseyRepresentationReconsidered2007}. The Level-1 conception renders effects as representations of their causes and causes as representations of their effects. If we think brains or robots represent the world in a richer sense than the one in which our heart rate represents our level of excitement or green leaves represent the past presence of sunlight, a more restrictive conception is needed.

\subsection{Level 2: Representations as Usable}

Going beyond Level 1, the concept of representation is often invoked to answer questions about how a system works or how it processes information about the world to achieve its goals. This is a different explanatory target than we discussed in the context of Level 1, and it requires a more refined notion of representation. At Level 2, representations are conceived of as internal states that \textit{potentially contribute to the successful behavior of an agent}. The ``agent'' here can be understood as any system with basic goals and actions, such as a rat or an autonomous car. For a state to be potentially used, it must at least (i) contain information related to the agent’s goal(s), and (ii) be formatted such that the information could feasibly be actionable. In other words, Level 2 adds constraints on Level 1 such that representations must be \textbf{useful} for goals and/or in a \textbf{usable format}.

One paradigmatic Level-2 analysis is quantifying whether a certain brain area or neural network layer ``represents'' $\mathbf{x}$ by the extent to which $\mathbf{x}$ can be linearly decoded from it \citep{alainUnderstandingIntermediateLayers2018,ivanovaLinearRegressionMapping2022}. Linear decoding analyses make an explicit assumption both about what information is relevant (the task the decoder is trained to solve) and what constitutes a usable format for that information (the decoder being linear). For example, neurons in primates’ infero-temporal cortex (IT) are said to represent certain semantic properties of visual objects in part because those semantic properties are considered relevant to primates’ goals and because they are linearly decodable from neural activity in IT \citep{hungFastReadoutObject2005}. Importantly, this kind of analysis cannot be used to say whether some internal state is actually used by an organism --- whether it has a causal effect on behavior --- but can only quantify the extent to which the goal-relevant information is present in a particular format and thus potentially usable by the system. At Level 2, this potential for use by the system is the key concept underlying ``representations'' \citep{ritchieDecodingBrainNeural2019}.

Another paradigmatic Level-2 analysis is Representational Similarity Analysis (RSA), which seeks to quantify the (dis)similarity of representations across different systems \citep{diedrichsenRepresentationalModelsCommon2017}. These analyses begin with a set of stimuli (e.g. images) and measurements of internal states for each of those stimuli (e.g. fMRI data or activations in a neural network hidden layer). They then quantify the extent to which internal states are (dis)similar to each other for different stimuli within each system for all pairs of stimuli. The similarity across systems is then quantified by comparing these pairwise-(dis)similarities. It has been shown that RSA and related analyses are equivalent to finding a ``simple'' mapping from one neural space to another, and this notion of simplicity in mapping across systems can be seen as encoding an assumption about what formats are usable within each system \citep{harveyWhatRepresentationalSimilarity2024}. For instance, pairwise Euclidean distances across stimuli within a system are sufficient to predict how well any dichotomy of the stimuli can be linearly decoded \citep{kriegeskortePeelingOnionBrain2019}, and linear decodability is a common proxy for usability \citep{alainUnderstandingIntermediateLayers2018,ivanovaLinearRegressionMapping2022,ritchieDecodingBrainNeural2019}. RSA addresses the extent to which two systems represent similar information in a similar format, on the assumption that these similar formats could be used by the system, thus incorporating the constraints of Level 2 (but not of Level 3).

\subsubsection{What constitutes usefulness?}

At Level 2, representations are understood as being useful, i.e. carrying some information that is relevant to the system’s goals. Different notions of goals can then lead to different notions of what is represented. We identified the following distinct notions of usefulness that contribute to a Level-2 conception of representations. 

\paragraph{Usefulness as task-relevance.} First and foremost, usefulness can be understood as task-relevance for a particular task. In a maze-navigation task, information about the maze structure, current location, and goal location are all useful, but details such as the wall texture may not be. Some analyses explicitly or implicitly require that a representation must carry information that is task-relevant, such as an analysis that seeks correlations between internal neural states and a rat’s location \citep{wikenheiserDecodingCognitiveMap2015} (cf. \cite{krakauerNeuroscienceNeedsBehavior2017,burnstonContentsVehiclesComplex2020,caoPuttingRepresentationsUse2022}). Such analyses often treat unaccounted-for variation in internal states as ``noise'' or a ``nuisance variable,'' even though it may reflect meaningful variations in task-irrelevant or unmeasured aspects of the environment \citep{musallSingletrialNeuralDynamics2019,manteContextdependentComputationRecurrent2013}. This is a natural decision. Representations link a system to its environment, helping us explain the system's behavior in terms of its relation to that environment \citep{richmondWhatTheoryNeural2025}. And what matters for both the system's behavior and our explanation of it are the features of the environment that are relevant for its tasks. 

On this notion of usefulness, a lot hinges on how we characterize the task. Navigating a maze is one task; finding food and water is another. Conceiving of a task in the second way might make wall texture relevant, e.g. if texture implies climbability, indicating options for finding food and water that involve leaving the maze. Different tasks, or different descriptions of them, can vary in generality; maze-navigation could be a special case of food-finding, which could itself be a special case of survival or maintaining homeostasis, etc. Some theoretical work takes a maximally task-agnostic approach, saying that useful representations are those that retain as much information as possible about the sensory input while reformatting that information to be more usable across a variety of tasks \citep{bengioRepresentationLearningReview2013}. A task-agnostic approach to usability is somewhat limited in application, however, in the absence of a mathematical formalization of collections of tasks. Some efforts toward such formalization have been taken up in work on multi-task learning \citep{caruanaMultitaskLearning1997, ben-davidExploitingTaskRelatedness2003}, but developing a comprehensive theory of task collections remains an open challenge.

What constitutes useful information can also vary based on how a system’s physical attributes interact with the space of tasks it encounters. For instance, if we want to create an embodied agent that can navigate rough terrain, we may give it legs with morphological properties that are suited to many different ground textures, obviating the need for centralized information-processing capabilities to represent that texture. Similarly, if a robot is fixed in one place, information about self-location may not be useful to represent \citep{robertsExamplesGibsonianAffordances2020, cisekAffordanceCompetitionHypothesis2008}. When an agent or system has no need to internalize or process certain kinds of information to achieve its goals, it may no longer be useful for that system to represent that information in a neural state, even though it is task-relevant.

Finally, tasks can vary in their time-bounds, and some representations may be better suited for immediate use or for future use. This is related to a distinction between inference (immediate) and learning (future). Solving a maze and learning to solve a maze are different tasks, the latter extending over a longer period of time and making it useful to track different information like the global structure of the maze. With tasks over longer time-scales, certain aspects of a representation can be useful not because they themselves contribute to solving the immediate task, but because they facilitate learning. Connecting this to some recent ideas in machine learning, the Lottery Ticket Hypothesis proposes that it is useful to start with a diverse array of structures to ensure that some of them contribute over the course of learning \citep{frankleLotteryTicketHypothesis2019}. In neuroscience, it has been argued that internal states should be variable or noisy to provide gradient signals for learning \citep{wuTemporalStructureMotor2014,lansdellNeuralSpikingCausal2023}.

\paragraph{Usefulness as optimality.} Some formal theories specify conditions on optimal performance of a task, or on optimal representation of information. There are diverse formalisms for optimality, leading to diverse notions of what makes certain information contents or formats better than others. On some conceptions, a maximally useful state is one that contains minimal and sufficient information to complete a task \citep{tishbyInformationBottleneckMethod1999,tishbyDeepLearningInformation2015,soattoVisualRepresentationsDefining2016,achilleEmergenceInvarianceDisentanglement2018}. On others, useful information would be whatever can be encoded in an efficient or sparse way \citep{simoncelliNaturalImageStatistics2001,bialekEfficientRepresentationDesign2007,olshausenSparseCodingIncomplete1997,barlowPossiblePrinciplesUnderlying1961,DayanAbbott2001,Chalk2018} (and see below on sparsity as a feature of representational format). And on others, it is most useful to represent not just contents, but probability distributions corresponding to them \citep{savageFoundationsStatistics1972,Kording2008,Zemel1998,Fiser2010,pougetProbabilisticBrainsKnowns2013}. For these notions of optimality, an internal state is more ``useful'' insofar as it is closer to the conditions for optimality for a given theory. These theories of optimality are often related to other notions of an agent’s high-level goals (e.g. metabolic efficiency supports survival), but in practice these theories are often decoupled from these high-level goals and function as qualities of ``useful'' representations in their own right. 

\paragraph{Usefulness as epistemic value.} Representations are often seen as central to understanding an agent's knowledge, especially in philosophy, suggesting a notion of epistemic usefulness. For example, an agent might believe that ``all birds fly,'' and this belief can be viewed as a kind of representation of how the world is, evaluable in terms of its truth or accuracy. If it is true or accurate (and certain other conditions are met) the representation may be taken to constitute knowledge \citep{fodorPsychosemanticsProblemMeaning1987,dretskeKnowledgeFlowInformation1986,millikanLanguageThoughtOther1987, sosaKnowledgePerspectiveSelected1991, zagzebskiVirtuesMindInquiry1996, haackEvidenceInquiryTowards1993}. With this conception of representation in mind, whether some internal state is ``useful'' depends on a number of factors. First, a state may or may not be truth-evaluable in the first place, and truth-evaluability is a precondition for representational content to be assessed against reality. A malformatted state or a state with nonsensical contents like ``all blergs floop`` has an undefined truth value. Second, provided that the first truth-evaluability criterion is met, contents that are more true or veridical are more epistemically useful than those that are false; a representation of the proposition ``all dogs are mammals'' is in this sense more useful than a representation of the proposition ``all birds fly.'' Finally, epistemic value is greater when a representation's content integrates with an agent’s existing knowledge \citep{haackEvidenceInquiryTowards1993}. For example, learning a new word in a familiar language is more useful than learning a new word in a foreign language that one doesn’t speak. This epistemic notion of usefulness can be contrasted with a pragmatic notion, more closely aligned with task-relevance, discussed above. Consider a representation of ``a predator is nearby''; for some creatures it may be pragmatically useful for such a representation to be systematically inaccurate in the direction of false positives, since fleeing from a non-predator is much less costly than failing to flee from a predator. Epistemic value can be thought of as a usefulness criterion that determines a representation's contribution to knowledge, independent of whichever actions the knowledge might support. 

\subsubsection{What constitutes a usable format?}

Goal-relevant information is only really useful if it is stored in a usable format, which is why we group them together at Level 2. A representation of food location might be very useful, but if it is formatted in such a way that the information is inaccessible, it cannot factor into computational or representational explanations of how a system works. With this in mind, we identified the following notions of ``usable format'' that contribute to a Level-2 conception of representations.

\paragraph{Usability as structure.}
A paradigm case of the Level-2 conception of representation is when we train a simple (often linear) decoder on neural data to test whether information is in principle available to be read out by a decoder of that kind \citep{alainUnderstandingIntermediateLayers2018}. Simple decoders can be justified pragmatically (fitting complex models can require prohibitive amounts of data) or as a proxy for the information that is available to a downstream neural network layer or downstream brain area \citep{ivanovaLinearRegressionMapping2022}, assuming that the brain can do linear but not nonlinear decoding. In neuroscience especially, the term ``explicit'' is often used to refer to representations whose information is accessible in a single neurally plausible stage of additional processing \citep{dicarloHowDoesBrain2012,hongExplicitInformationCategoryorthogonal2016,kriegeskortePeelingOnionBrain2019}. On these assumptions, linearly decodable information is stored in a usable format.

For neural systems, linear decoding may be too restrictive of an assumption in the sense that single neurons can do nonlinear things \citep{barlowPossiblePrinciplesUnderlying1961,poiraziMemoryCapacityLinear1999,carandiniNormalizationCanonicalNeural2012,goldenConjecturesRegardingNonlinear2016,larkumAreDendritesConceptually2022}, or it might be not restrictive enough in the sense that neural connections are typically sparse, so off-axis rotations of an internal state may not always be considered equivalent representations \citep{kriegeskortePeelingOnionBrain2019,khoslaSoftMatchingDistance2024}. Given a sufficiently complicated and highly specialized nonlinear decoder to decode information, one could say that the retina represents object identity. And indeed such a specialized decoder certainly exists in the average person’s brain. Interpreting ``usable format'' in this way could mean that an individual’s retina represents object identity for that individual, but not for other computationally-limited third-party eavesdroppers, just as an encrypted file on a computer could be said to represent its original contents only for those with the decryption key and the means to decrypt it. When the concept of usable format is invoked in these contexts it is always relative to assumptions about the capacities of the decoder or user of the representation.

In both machine learning and neuroscience, ``disentangled'' representations are often cited as a particularly usable format \citep{dicarloUntanglingInvariantObject2007,bengioRepresentationLearningReview2013,higginsDefinitionDisentangledRepresentations2018,achilleEmergenceInvarianceDisentanglement2018,achilleLifelongDisentangledRepresentation2018}. However, the term ``disentanglement'' is used in the literature to refer to different things, and many open questions remain regarding how disentangled representations (however defined) would lead to improved performance. This is an active area of research in theoretical machine learning, and progress is being made in these directions \citep{locatelloSoberLookUnsupervised2020,vansteenkisteAreDisentangledRepresentations2019,monteroRoleDisentanglementGeneralization2021,lipplWhenDoesCompositional2024,achilleEmergenceInvarianceDisentanglement2018,higginsDefinitionDisentangledRepresentations2018,hyvarinenNonlinearIndependentComponent2023,scholkopfCausalRepresentationLearning2021}. Recent theoretical work has also taken the tack of optimizing models to directly maximize the ``decodable'' information \citep{duboisLearningOptimalRepresentations2020}.

Decodability and disentanglement are both special cases of \textit{structural constraints} on representations, such that for a representation to be usable is for it to be structured in such a way that can be leveraged by downstream processes, for the functions those processes perform (e.g., object-detection). Some kinds of downstream functions may require specially formatted information. For instance, Language of Thought theories propose operations that can exploit only representations formatted according to specific syntactic rules \citep{fodorLanguageThought1975,aydedeLanguageThoughtHypothesis2010,quilty-dunnBestGameTown2022}.

A related notion of representational structure concerns a representation's relation, not just to the operations performed on it, but to \textit{what it represents}. Representations may be structured so that they preserve certain relationships in the represented object, so that those relationships can influence later processing. For instance, one might expect to find representations involved in visualizing a scene or imagining an object-interaction would preserve many spatial properties of actual places and objects. This property of preserving relations-in-the-world as relations-in-the-internal-state is known as ``iconic'' format in the philosophy literature \citep{kosslynMediumMessageMental1981,shepardMentalRotationThreedimensional1971,burgeIconicRepresentationMaps2018} and ``equivariance'' in the machine learning literature \citep{cohenLearningIrreducibleRepresentations2014,cohenGroupEquivariantConvolutional2016,higginsDefinitionDisentangledRepresentations2018,keurtiHomomorphismAutoEncoderLearning2023}. As we discussed in Level 1, connecting an internal state mechanistically to a source of sensory data provides one in-road into the philosophical concept of ``grounding'' representations in the world. Iconic or equivariant representations provide another potential link to grounding: if an internal state is a grounded visual representation, one might expect certain iconic properties of images, i.e. certain equivariance properties as the visual input or imagined visual scene are varied.

\paragraph{Usability as modality-independence or invariance.} Like equivariance and disentanglement, the usability of representations may depend on their level of invariance to input changes, and even invariance to different input modalities. An internal state with the content ``there is a cat'' may be considered more usable if it encodes this information regardless of the angle from which the cat is seen, or in fact whether the cat is seen, felt, or heard \citep{fukushimaNeocognitronSelforganizingNeural1980,rustAmbiguityInvarianceTwo2010,quirogaGnosticCells21st2013}. Highly invariant representations can be more useful in the sense that they facilitate efficient downstream learning and processing, and they can be more usable in the sense that they may serve as input to a diverse array of downstream functions, and those functions may be simpler in virtue of nuisance information being stripped away. Some recent work in machine learning has used cross-modality invariance as a learning objective, and this is emerging as a powerful technique for general-purpose representation-learning \citep{bachmanLearningRepresentationsMaximizing2019,tianContrastiveMultiviewCoding2020a,daunhawerIdentifiabilityResultsMultimodal2023,huangComparisonMultimodalSinglemodal2024}.

\paragraph{Usability as availability.} Another notion of usable format concerns a representation's place in the functional structure of a system, and the extent to which activity in one area is able to modulate activity in another. If an internal state is localized to area A, with no ability to influence area B (or unable to do so fast enough or in a way that preserves information or etc.), then the state is not usable by processes in area B. Merely formatting information in one area for hypothetical decoding may not be sufficient to call it a representation if that information is not then made available through projections to other brain areas or other parts of the system --- the so-called ``representation'' would have no role in the generating the organism's behavior, or in any explanation thereof. Considering the availability of information to downstream systems begins to introduce Level-3 concepts, but a sharp distinction between the levels is not our primary goal in this paper.

\subsubsection{What does Level 2 let us do? What questions does it answer?}

\begin{figure}[bt]
    \centering
    \includegraphics[width=0.99\textwidth]{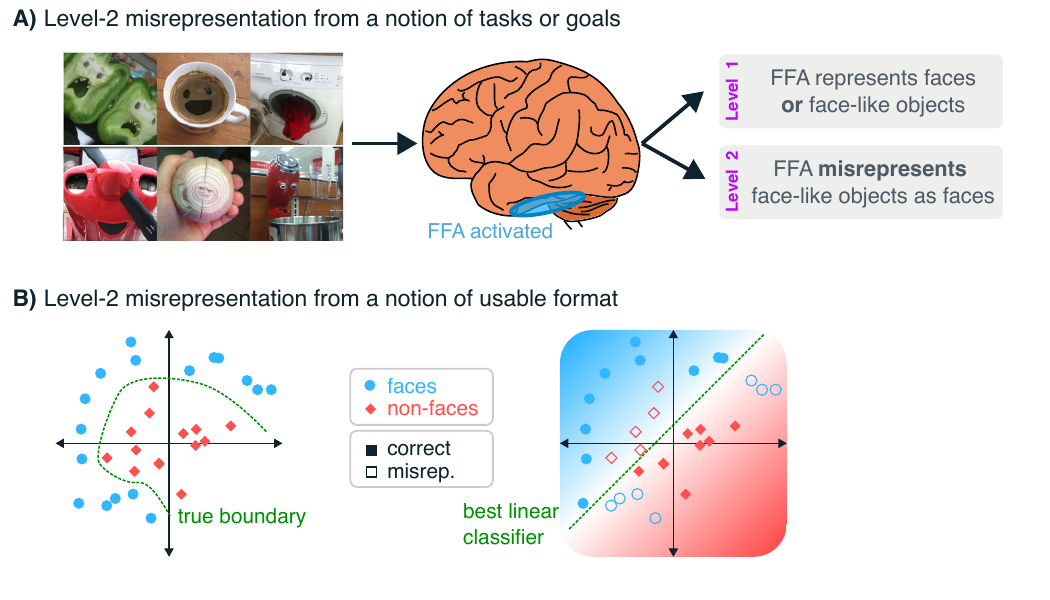}
    \caption{Illustration of two ways that Level 2 allows for a notion of misrepresentation (and Level 1 does not). \textbf{A) An account of misrepresentation from assumptions about the task.} For the sake of argument, say that these face-like objects (from \citep{wardleRapidDynamicProcessing2020}) elicit neural activity in the Fusiform Face Area (FFA) indistinguishable from the neural response to actual faces. (This is just an illustrative argument. Note that \citep{wardleRapidDynamicProcessing2020} do report differences in FFA; see the original study for details.) At Level 1, one interprets FFA as representing the disjunction of faces or face-like objects, since it contains reliable information about both. By hypothesizing that faces are more ecologically relevant than face-like objects, one introduces a notion of what constitutes \textbf{usefulness}, thus moving to Level 2 and allowing a notion of misrepresentation of face-like objects as faces. \textbf{B) An account of misrepresentation from assumptions about usable format.} On the left, we imagine a space of patterns of neural activity responding to both faces and face-like objects. The two classes may in principle be well-separated, but not in a way that is usable by a particular family of limited-capacity decoder. On the right, constraining the notion of \textbf{usable format} to information that is linearly decodable, the in principle best decoder makes mistakes which may be considered misrepresentations.}
    \label{fig:faces-misrepresentation}
\end{figure}

First, Level 2 enables ``\textbf{how possibly}'' answers to questions about how systems function using internal representations \citep{verreault-julienHowCouldModels2019, bokulichHowTigerBush2014,Brainard2020-BRAHTE-2}. That is, if an organism or computer system exhibits some behavior that putatively involves representations, then identifying internal states which carry useful or task-relevant information formatted in a usable way is a major step towards understanding how the system functions. Level 2 thus begins to ground representational hypotheses --- like what is represented where and how --- in data.

A typical hypothesis at Level 2 is something like ``[relevant information] is represented in [layer or area] in [format]''. Here are two concrete examples:
\begin{itemize}
    \item Face/nonface, but not individual face identity, is represented in the fusiform face area (FFA) in such a way that it can be linearly decoded from fMRI signals. It is likely that anterior inferotemporal cortex (aIT) is more involved in representing face identity in virtue of identity being more readily decoded from aIT than from FFA \citep{kriegeskorteIndividualFacesElicit2007}.
    \item Object pose (and other nuisance variables) are represented by intermediate layers of a CNN trained only on categorization, such that nuisance variables can be linearly decoded from intermediate layers but not from input or output layers \citep{patelProbabilisticTheoryDeep2015}.
\end{itemize}

Note that the second example --- linearly decoding nuisance variables --- could be seen as a case of ``usable but not useful'' information for a narrow task-specific definition of usefulness. For a broader notion of usefulness that encompasses other tasks, the ability to linearly decode ``nuisance'' variables may still be seen as a case of useful information.

Importantly, these hypotheses about what is represented where and how are falsifiable by constructing a decoder of the specified class and seeing if it can in fact extract the hypothesized information from the specified state. However, such experiments test a conjunction of hypotheses about both relevant information and format. Thus, falsifying a Level-2 hypothesis about what is represented and how can leave open the alternative that some information may be present but in a different format. For example, failure to decode the hypothesized information from fMRI BOLD signal does not imply that this information is not present, as it could be more localized or more transient than is detectable with fMRI \citep{logothetisWhatWeCan2008}.

In allowing that the usefulness of an internal state may be determined by its contribution to learning (rather than by applicability to an ongoing task), Level 2 supports ``how possibly'' accounts of an agent’s ability to imagine or plan. These are examples where carrying information about the state of the world is not of primary importance, and also examples of what philosophers have called ``representation-hungry'' abilities, which involve sensitivity to distal, non-existent, or highly abstract entities \citep{clarkDoingRepresenting1994,degenaarRepresentationhungerReconsidered2014,schlosserEmbodiedCognitionTemporally2018}. Behaviors that deal only with aspects of the environment for which information is available to its sensory apparatus are at least potentially amenable to explanation in terms of the causal dynamics of the agent-environment system, without representation playing an important role. Representation-hungry abilities, on the other hand, seem relatively inexplicable except in terms of internal states whose function depends on what they represent. For instance, when place cells are interpreted as representations of imagined trajectories through a maze which a mouse has not experienced \citep{dragoiPreplayFuturePlace2011}, they can still be understood as useful toward a plausible future task. The imagined trajectories contain information relevant to possible future tasks and they facilitate learning with respect to those tasks \citep{buesingWouldaCouldaShoulda2018,hamrickAnaloguesMentalSimulation2019,jannerWhenTrustYour2019}. This is an important point: Level-2 representations need not carry information about the present state of the world, and this begins to decouple Level 2 from Level 1.

Similarly, Level 2 allows for a concept of misrepresentation and ``how possibly'' representational explanations of why agents sometimes make errors. For instance, we can consider a ``how possibly'' explanation of the way people often see electric outlets as faces (called pareidolia). Supposing FFA is hypothesized to represent the presence of a face, one might hope to identify circumstances wherein FFA systematically fails to carry the face-presence information that it is its job to represent --- cases where FFA responds as it characteristically does to faces, but when there is no face present. At Level 1, such an internal state could be thought of as a well-formed representation of a \textit{disjunction}: objects that are either faces or outlets. If, instead, a representation is understood with respect to a relevant task, like ``identify other humans,'' then there is a role for a notion of mis-representation to mark cases where, e.g., face-representations of electrical outlets fail to serve that task (Figure \ref{fig:faces-misrepresentation}A) \citep{wardleRapidDynamicProcessing2020, richmondWhatTheoryNeural2025}. A similar story can be told about representational format: if a system is hypothesized to use some limited-capacity decoder, then that system can be understood as misrepresenting information when an edge-case instance is handled incorrectly by that particular decoder (Figure \ref{fig:faces-misrepresentation}B). 

A Level-2 conception can also figure in preliminary hypotheses about the epistemic status of an agent. For instance, one can provide partial support for a claim like ``this animal knows that P,'' where P is some proposition, by showing that P is encoded by an internal state that is truth-evaluable and carried in a usable format. Notably, philosophical accounts of knowledge generally require more than a true and relevant representation; they require something about how one comes to have a representation, in particular whether it is justified or is a product of the agent's intellectual virtues \citep{gettierJustifiedTrueBelief1963,sosaKnowledgePerspectiveSelected1991,zagzebskiVirtuesMindInquiry1996}. But still, Level 2 use can be understood in relation to epistemic achievement, providing a coherent starting point for asking what kinds of representations count towards an agent’s knowledge. 

Level 2 also provides a normative account of the kinds of computations that need to happen upstream of a given representation. If information about objects is highly entangled in early stages of visual processing but linearly decodable by later stages, one can explain the role of the intervening layers as reformatting the unusable information early on into a bona-fide Level 2 representation in those later stages \citep{dicarloUntanglingInvariantObject2007}. In contrast, applying the logic of Level 1, information processing or reformatting can at best preserve information and will in general destroy information (a consequence of the Information Processing Inequality \citep{coverElementsInformationTheory2012}). At Level 2, the added constraint that information must be usable implies that representational content can be gained through an increase in ``usable'' information \citep{xuTheoryUsableInformation2020,kriegeskorteInterpretingEncodingDecoding2019,kleinmanUsableInformationEvolution2021}. This, in turn, provides a normative account of neural information processing in terms of the usability of the representations that it produces \citep{langeDeepNetworksPaths2023}. Notably, this account of information-processing as changing representational formats transcends both neuroscience and ML.

Researchers who operate at Level 2 often take a modular approach, decomposing systems into subsystems that represent distinct contents and serve distinct goals. This kind of assumption is present in the FFA example above, as FFA is a localized region of the brain purported to carry out the specialized goal of face identification. Likewise in the case of the object-recognizing CNN above, each layer of the CNN is a localized region of the broader network, and the hypothesis of \citet{patelProbabilisticTheoryDeep2015} assigned the intermediate layers the sub-goal of disentangling variables along the way towards the whole network’s goal of classification. Modular decomposition of representational contents and goals is a convenient and sometimes-justifiable assumption for researchers seeking to reverse-engineer the representations in brains and neural networks. We say that Level 2 ``allows'' a modular approach in the sense that modularity itself can be seen as a design-principle for greater usability (related to disentanglement), and also in the sense that ``usefulness'' takes on a more fine-grained meaning when considering sub-goals that parts of the system may have. Level 2 thus allows for ``how possibly'' accounts of the way representational subsystems compose the whole-system’s behavior (FFA detects a face, aIT recognizes the face as a friend, and motor cortex orchestrates waving hello), even without yet specifying interactions between them.

Importantly, the modular assumption can be taken too far. First, groups of neurons may be involved in many functions, and many groups of neurons may be involved in single functions \citep{andersonPhrenologyNeuralReuse2014,fusiWhyNeuronsMix2016,pessoaEntangledBrain2022,nobleTipIcebergCall2024}. Second, some putatively representational capacities may emerge from widely distributed neural dynamics --- distributed in both space and time \citep{barackMentalKinematicsDynamics2021,burnstonGettingAtomismFunctional2021}. However, while dynamics like this might frustrate the modular approach, it does not cause trouble for our broader framework. Instead, the dynamics perspective would require that ``internal state'' and ``usability'' be recast in dynamical systems terms.

Finally, Level 2 can also support the design of new systems that use the information present in the system under study. For instance, a Level-2 perspective is apt in pre-training and fine-tuning in machine learning, and to a limited extent also in building  brain-computer interfaces (BCIs). Creating a BCI requires that useful or relevant information is available in a decodable way, whether or not it is currently used in behavior. Thus the process typically begins with identifying a plausible (linear) decoder from information already present in the neural signal \citep{carmenaLearningControlBrain2003}. For instance, neural activity that correlates with intended movement without driving any muscle activity could be an ideal candidate for controlling an external device such as a robotic arm. However, research on rhesus macaques suggests that the neural activity patterns that can be effectively mapped to control a BCI may be largely constrained to those patterns that are already well established \citep{sadtlerNeuralConstraintsLearning2014, golubLearningNeuralReassociation2018}. The fact that these patterns are well established presumably depends on their having significant downstream effects --- their being already in use --- which would suggest not just usability, but actual use has relevance for BCI design. In contrast, downstream effects can be ignored when it comes to training a machine learning model in one domain and transferring or fine-tuning the model for use in another domain. This requires only that the initial model contains the right information in a usable format, independent of whether or how that information was used in the original model.

\subsubsection{What does Level 2 miss? What questions does it fail to answer?}

First, Level 2 is dependent on the assumptions and hypotheses of the scientist or engineer about what makes a representation useful and what makes it usable. The tasks that are used to probe a system require assumptions about what is ecologically relevant. Disagreements about the aptness of a particular decoder or about the ecological goals of a system might seem to constitute disagreements about what is represented, where, and how. But at Level 2, given a reasonably broad notion of usability, such different representations are not incompatible but reflect different conceptions of, or ways of framing, an organism's tasks, resulting in differences in what information is relevant to the scientist explaining those tasks. However, Level 2 does not present a way of \textit{settling} disagreements about which use-cases, tasks, or task conceptions should inform our understanding of the system.

Similarly, when it comes to determining what is ``usable,'' taking a particular type of decoder to probe usability requires an assumption about what kinds of things are in principle decodable by the system. For instance, common practice is to apply a linear decoder to both biological and artificial systems, but of course this one-size-fits-all approach can be inflexible, not allowing for different representational formats that may be used by different systems \citep{ivanovaLinearRegressionMapping2022}.

Finally, there are issues of correlation versus causation at Level 2. An internal state may contain in principle task-relevant information in an in-principle decodable format, but this internal state could be causally inert in the system in which it occurs, or it could be interpreted by downstream systems in a way that is inconsistent with its hypothesized representational role. Since potential use does not strictly entail actual use, those states cannot factor into the kinds of representational explanation of how a system actually functions and performs its tasks that are typically sought by cognitive scientists \citep{decharmsNeuralRepresentationCortical2000,ritchieDecodingBrainNeural2019,mekikCognitiveScienceNutshell2022}. This is especially challenging in a system as complex as the brain, and indeed it has been demonstrated empirically that not all ``decodable'' information is causally efficacious \citep{grootswagersFindingDecodableInformation2018}. Just as Level 2 is motivated in part to by expanding on the Level-1 notion of representation as ``mere information,'' the next step to Level 3 is motivated in part by expanding on the Level-2 notion of representation as ``mere potential'' for use.

\subsection{Level 3: Representations as Used}

Where Level 2 incorporated notions of representations containing useful information and formatted in a potentially usable way, Level 3 will incorporate a notion of representations being actually used. Actual use has to do with the causal role an internal state has on downstream functions or behavior. As before, Level 3 is not necessarily ``better'' than Level 2, but it asks different questions and serves different explanatory goals.

To illustrate, consider a motion direction estimation task in which a subject views a field of moving dots whose overall movement trends in one direction, and the subject’s task is to judge which direction that is. Parts of the subject’s nervous system may carry information about many different motion directions, such as the directions of individual dots and clusters of dots. All of the direction information is potentially useful, so all of these internal states are representations at Level 2. However, it could be that only a subset of these internal, direction-indicating states causally contribute to a final decision about which direction the field of dots is moving on average. Researchers who are specifically interested in the states that make a difference to the observed perceptual judgment would identify only the causally efficacious states as representations. Indeed, a crucial piece of the well-known story that mediotemporal cortex (MT) represents motion direction is that causal interventions on MT influences a monkey’s decisions on tasks like this one \citep{nicholsMiddleTemporalVisual2002}. Parts of the nervous system for which intervention has no discernible effect, whether their activity carries potentially usable information about motion direction or not, are not seen as representations of motion direction at Level 3.

Research that conceives of representations at Level 3 understands the content of representations to have more to do with its normal downstream effects than with its upstream causes or use-potential \citep{andersonContentActionGuidance2008}. Especially in philosophy and sometimes in neuroscience, a focus on explaining observed, goal-directed action supports thinking about representations primarily in terms of their causal roles in supporting that behavior. These approaches are particularly concerned to distinguish between states that merely correlate with task-relevant variables and those that contribute significantly to adaptive behavior. As such, in order to determine that an internal state of interest is genuinely a representation, emphasis is placed on precisely specifying the behavior that the internal state supports, and sometimes on establishing that the proposed causal role of the information-carrying state was supported by a historical process of selection \citep{millikanLanguageThoughtOther1987,dretskeExplainingBehaviorReasons1991,sheaRepresentationCognitiveScience2018}.

\subsubsection{What constitutes actual use?}

First, one can think of ``using'' an internal state in terms of its causal influence on outward behavior. This is perhaps the most straightforward interpretation of ``use'' in Level-3 thinking in the sense that it is relatively easy to measure and it requires few assumptions about other internal information-processing structures. This interpretation is suggested in the examples just considered, and has a rich history in neuroscience and philosophy. Following \citet{parkerSenseSingleNeuron1998}, in order to justify calling some neural activity a perceptual representation, neuroscientists sometimes require that manipulation of the neural activity should affect perceptual judgments, and prevention of the neural activity should impair perceptual abilities, as measured by behavioral reports. A substantial literature in philosophy argues that representational content is grounded in causal relationships and functional roles, including etiological functions \citep{stampeCausalTheoryLinguistic1977,macdonaldTeleosemantics2006,dretskeExplainingBehaviorReasons1991}. Under this conception, internal states qualify as representations only if they have a demonstrable causal role in shaping an agent or system’s outward actions.

Second, one can think of an internal state being ``used'' in terms of its causal influence on other downstream subsystems. That is, internal states in one part of the system may play representational roles by affecting other parts of the system, even without a discernible manifestation in behavior. For instance, consider neurons in the prefrontal cortex responsible for working memory. These may carry task-relevant information in their sustained activity patterns without any observable effect, in particular if what is remembered in the short term turns out not to be actionable. Some representations can in this way be understood as being used by the working-memory system, even in the absence of any discernible or immediate impact on behavior. This second notion of ``internal use'' depends on knowledge or assumptions about modular design, such as distinguishing which internal states are involved in ``perception'' and which are involved in ``memory'' or ``decision-making.''

The line between Level 2 and Level 3 is sometimes fuzzy and may be a matter of perspective. Earlier in Level 2, we suggested that one’s retina could be seen as representing objects in a ``usable format,'' taking ``usable'' to mean with respect to the particular individual’s downstream visual system. This arguably pushes into Level-3 territory by considering an extant rather than hypothetical downstream architecture. Many researchers who work at Level 2 may also be thinking in terms of Level 3, under the reasonable assumption that information that is present, relevant, and formatted for easy decoding is probably used by the system \citep{kriegeskortePeelingOnionBrain2019}, though this assumption is not always justified \citep{ritchieDecodingBrainNeural2019}. But note also that Level-3 investigations might \textit{set aside} considerations of format and precise content. To show that a representation is used in some task is not necessarily to specify the format that enables that use, or even the specific task-relevant content of that representation. Consider the investigations of frog prey detection that revealed the neural representations of frog prey that support prey detection \citep{lettvinWhatFrogsEye1959}, but were \textit{not} able to determine the format of the representations or their content --- aspects of the representations that are still debated today \citep{millikanTeleosemanticsFrogs2024}. We reiterate that our aim here is not to endorse one or another representational theory, but to highlight that potential and actual use (Level 2 and Level 3) are distinct ways of thinking about neural representations.

\subsubsection{What sorts of analysis or insights does Level 3 allow?}

First, Level 3 enables ``\textbf{how actually}'' explanations of a system's functional components, beyond ``how possibly'' explanations \citep{strevensDepthAccountScientific2011,bokulichHowTigerBush2014,verreault-julienHowCouldModels2019}. A ``how actually'' explanation invokes actual, measured downstream systems that are causally influenced by some internal state, and often perturbation studies to empirically validate how the internal state is used. This focus on causal influence allows researchers to connect internal states to observed behavioral capacities. For example, while information-theoretic analyses might suggest the fusiform face area (FFA) carries information about faces at Level 2, establishing its representational status at Level 3 requires evidence that this activity causally influences face-related behaviors. Such evidence comes from intervention studies showing that electrical stimulation of FFA produces face-like percepts \citep{schalkFacephenesRainbowsCausal2017} and that FFA lesions specifically impair face recognition while sparing other visual abilities \citep{bartonLesionsFusiformFace2002}. Similarly, while large language models may contain internal states that correlate with semantic concepts (Level 2), demonstrating these are genuine semantic representations at Level 3 requires showing that intervening on these states predictably affects model outputs \citep{brickenMonosemanticityDecomposingLanguage2023,templetonScalingMonosemanticityExtracting2024,lindsey2025biology}.

It might seem like these ``how actually'' explanations rely on direct interventions on internal states, but it is also sometimes possible to infer how those states are used from variations in \textit{stimuli}, and the changes those stimuli cause in a system's behavior. For example, consider studies purporting to show that prey-capture behavior in frogs does not change when their prey --- worms --- are replaced with cardboard cutouts sharing only the general shape and overall movement patterns of worms. A natural inference is that the task of prey detection only uses representations of features that are shared between the cardboard cutouts and the actual worms. Likewise, if prey capture behavior \textit{had} changed, this would be some evidence that representations of the missing features \textit{were} being used in the processes responsible for prey capture. (However, as with any studies using unnatural stimuli, these inferences would be undermined if the frogs operated by different internal mechanisms in these unnatural contexts \citep{gibsonTheoryDirectVisual2002}.) Similar inferences are possible from examining \textit{internal} activity in response to different stimuli, but fall under the discussion in Level 2: if internal activity changes (or does not change) depending on stimuli, that tells us there is (or is not) useful information, but not whether it is used in the production of behavior.

Second, Level 3 provides an alternative way to think about misrepresentations to that in Level 2 --- one that pertains directly to particular, observed errors, rather than how general types of error may occur. When some goal-directed system makes a mistake, one may investigate it to discover which (mis)representation is responsible for the (mis)behavior. When a human or robot hand grasps at thin air as though a solid object were there, a reasonable conjecture is that this is due to some internal state having had a downstream effect as if there were an object there. When one is trying to explain or fix the error, the internal states of interest are those that have this ``as if'' effect – those that led to the behavior that would have been apt if an object were within reach. Visual hallucinations are a clear example of this, and may be understood as internal states causing behavioral responses appropriate to non-existent sources of visual information. Thus, at Level 3, a misrepresentation is identified not merely by a mismatch between usable information and task success (as in Level 2), but by its causal influence in failure at the task, highlighting internal states that drive errant actions \citep[cf.][]{richmondWhatTheoryNeural2025}.

Contrasting Level 2 with Level 3 also suggests ``misuse'' as another kind of misrepresentation. An internal state might in principle contain perfectly well-formatted and task-relevant information, but it is up to the particular system in which it occurs to extract and use that information appropriately. For example, we can return to the example of prism glasses introduced earlier, which distort one’s perception of space and can lead to errant behaviors like grasping to the side of an object instead of grasping the object. For someone wearing prism glasses, information about the visual world is relatively well-preserved throughout the visual system. There is, in principle, information about the location of the object that is both relevant and usable, but there is a disconnect in how that information leads to a reach-and-grasp behavior. Explaining errors in the internal or external behavior of a system requires understanding not just how a particular state is susceptible to errant variability, but also how the state actually influences things downstream \citep{beckNotNoisyJust2012}. That is, a more nuanced conception of representations at Level 3 asks not just \emph{whether} a state is used to influence behavior, but \emph{how} it is used or misused.

Because representations can be repurposed beyond their original use-case, the line between use and misuse, or between representation and misrepresentation, can be somewhat indefinite or can change over time. A familiar example is using space to represent time, as when pointing to a date on a timeline to schedule a meeting. In such a case, a representation (of spatial information) takes on a new representational role (of temporal relations), departing from its original function. Similarly, neural activity in the visual word form area, which may have evolved to represent faces and objects, may be repurposed or ``recycled'' to carry content related to text, serving modern linguistic abilities \citep{dehaeneUniqueRoleVisual2011}. In artificial systems, representations developed through training for one task may be repurposed through training the network on a new task \citep{tanSurveyDeepTransfer2018}. Such ``repurposing'' of a representation for a new goal or task is not necessarily a kind of ``misuse'' or ``misrepresentation.'' Instead, representations can acquire new representational roles in this way, and so Level 3 incorporates a degree of pluralism reminiscent of Level 2: there may be multiple answers to what a given internal state is used for, and thus multiple legitimate perspectives on \emph{how} a representation is used, or whether it is misused in a particular instance. A difference from Level 2 is that this pluralism concerns not just what information carried by an internal state is potentially useful, but the multiple causal roles that the state could play.

Some recent developments in machine learning are also based in Level-3 thinking about representations. A significant body of work on deep neural networks has focused on methods that encourage properties like modularity, disentanglement, and equivariance, which are thought to count toward usability in principle (Level 2) \citep{ridgewayLearningDeepDisentangled2018,eastwoodFrameworkQuantitativeEvaluation2018,bengioRepresentationLearningReview2013,higginsDefinitionDisentangledRepresentations2018}. However, recent studies have questioned how important such properties are to performance. For instance, \citet{monteroRoleDisentanglementGeneralization2021} showed that greater disentanglement does not necessarily lead to improved combinatorial generalization. We view this as an illustrative example of the disconnect between potential use at Level 2 and actual use at Level 3: it may be the case that ``disentangled'' representations are necessary for combinatorial generalization, but they are evidently insufficient when actually tested. 

A similar story can be told about RSA, which we argued earlier is a Level-2 analysis because it uses correlational methods that do not consider downstream effects in comparing internal states across networks \citep{kornblithSimilarityNeuralNetwork2019}. Recent ML work on ``neural stitching'' is an example of a Level-3 analysis comparing representations across systems. Neural stitching compares internal states across models by surgically ``stitching'' activity from one model into another, then running the result through the latter model’s downstream layers. To the extent that this results in minimal degradation in the models’ downstream performance, we conclude that the systems had compatible representations \citep{lencUnderstandingImageRepresentations2015,bansalRevisitingModelStitching2021,csiszarikSimilarityMatchingNeural2021}. Thus, actual downstream use can also play a role in how we think about comparing representations across different systems.

\subsubsection{What limitations or drawbacks does Level 3 have?}

One significant limitation of the Level-3 approach is the inherent difficulty in establishing causality, particularly in complex, imperfectly observed systems like the brain. Since biological organisms typically have many neurons and synaptic connections, and the neurons themselves are often complicated functions of multiple types of input, it is very challenging to isolate distinctive internal states in terms of their exact causal role. Experimental interventions, such as inactivating or stimulating nerve activity, often affect whole regions of the nervous system rather than specific circuits or cells, which may not provide the granularity needed for proper causal analysis. Further, methods that attempt to infer causal relationships using time delays (e.g., Granger causality) can be unreliable given the complex temporal dynamics of neural processes. As we suggested earlier, some representational hypotheses can be falsified at Level 2, e.g. by showing that a particular layer of a neural network does not in fact contain easily-decodable information for some task. However, even when easily-decodable information is present, further (often causal) tests are  necessary to establish a ``representational'' role in the sense of Level 3 \citep{grootswagersFindingDecodableInformation2018, bakerThreeAspectsRepresentation2022}.

Another limitation of Level 3 is that merely establishing that a state plays a causal role in the production of behavior does not necessarily provide a single clean answer for how we should think about the representational format or content. Rather, a particular causal profile must be established to show that an internal state is used in the way entailed by the hypothesized content, adding to the difficulty arising from the limited ability to make precise interventions on representations. As we discussed above, we have known for decades that the frog's optic tectum responds to moving dark spots and mediates the frog's prey capture response \citep{lettvinWhatFrogsEye1959}. But even if this constitutes a discovery of a representation, it does not yet show us what \textit{format} that representation has (at least without much more precise interventions). Nor does it settle the \textit{content} of that representation --- whether it is more like ``frog food'' or ``small moving objects'' --- a debate which continues to this day \citep{millikanTeleosemanticsFrogs2024}. Similarly, a stroke or internally generated noise might affect the prey capture response, but not in virtue of having goal-relevant content, whereas a hallucinated worm generated by stimulating just the right neurons \textit{would} have a content-driven effect, and the distinction between these involves more than the bare fact of causal mediation. This means that, like Level 2, conclusions about representational format and representational content at Level 3 are theory-laden, not theory-neutral --- they depend on hypotheses about the system's goals and the contributions of various parts toward those goals. Thus, when thinking at Level 3, researchers often must sacrifice specificity about the format and content of a representation in recognition of the limited precision with which causality can be measured.

Another key limitation of Level 3 is that it struggles to capture internal states that are important for learning or developing future capabilities but do not yet play a causal role in behavior. Consider someone who has memorized a phrase in a foreign language phonetically, without knowing its meaning. Initially, this internally stored pattern does not significantly influence their behavior when they encounter the spoken language --- they cannot use it to understand or respond appropriately to others. As they gradually learn the language, the memorized phrase eventually takes on a new causal role, enabling them to respond in ways that are sensitive to what the phrase means. To understand how such stored patterns contribute to the learning process, a Level-2 approach is more suitable, since it encompasses potentially useful representations before they play any causal role in behavior. More broadly, the Level-3 requirement of causal influence on behavior rules out states whose useful information will only become efficacious once new abilities or new ways of encoding information become available to the system. In many domains besides language, it can be productive to assess the representational content of internal states that do not shape behavior now, but will shape behavior under common circumstances of development.

\section{Open Questions \& Discussion}

\subsection{What level are you?}

\begin{figure}
    \centering
    \includegraphics[scale=.75]{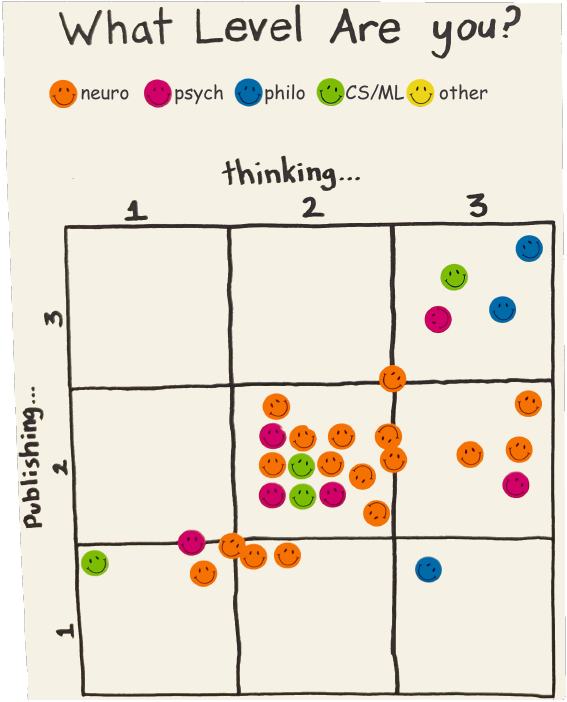}
    \caption{Results from a live poll conducted at the poster session of CCN 2024 (\url{https://2024.ccneuro.org/}). The data are informal, preliminary, and reflect a particular audience, but even within this small sample it illustrates a diversity of opinions on ``representations'' and some distinction between the level at which participants ``think'' and ``publish'' about representations.}
    \label{fig:poll}
\end{figure}

At CCN 2024, we conducted an informal live poll of attendees who visited our poster. After explaining the three Levels, we asked each visitor to locate themselves in a two-axis space (Figure \ref{fig:poll}). One axis asks participants: when they \emph{think about} representations, do they think primarily about information (Level 1), potential usefulness or usability (Level 2), or downstream causal effects (Level 3)? The second axis asks participants the same question about the way they operationally use the term ``representation'' in their published work. These axes are meant to capture the possible difference between the views a researcher might hold about representation in general from the claims about representation they are able to support to the level required for publication.  Participants self-reported their level by placing a sticker on the grid, using the color of the sticker to indicate which field they considered themselves to be a part of.

While informal and preliminary, the results are nonetheless interesting. Most notably, we see that participants gave a range of answers, indicating the diversity of existing ideas and approaches to studying neural representations. Further, no participants indicated that they \emph{publish} at a higher level than they \emph{think}. As we have stressed throughout, our three-level framework is not intended to carry a value judgment; there is important work being done at all three levels, and a study using Level 3 concepts is not necessarily stronger than a study using Level-2 concepts. Still, the fact that some researchers \emph{think} at Level 2 but \emph{publish} at Level 1 or \emph{think} at Level 3 but \emph{publish} at Level 2 --- but never the other way around --- suggests that the levels that we identified are meaningful.

\subsection{Who is the user?}

\begin{figure}[bt]
    \centering\includegraphics[width=0.99\textwidth]{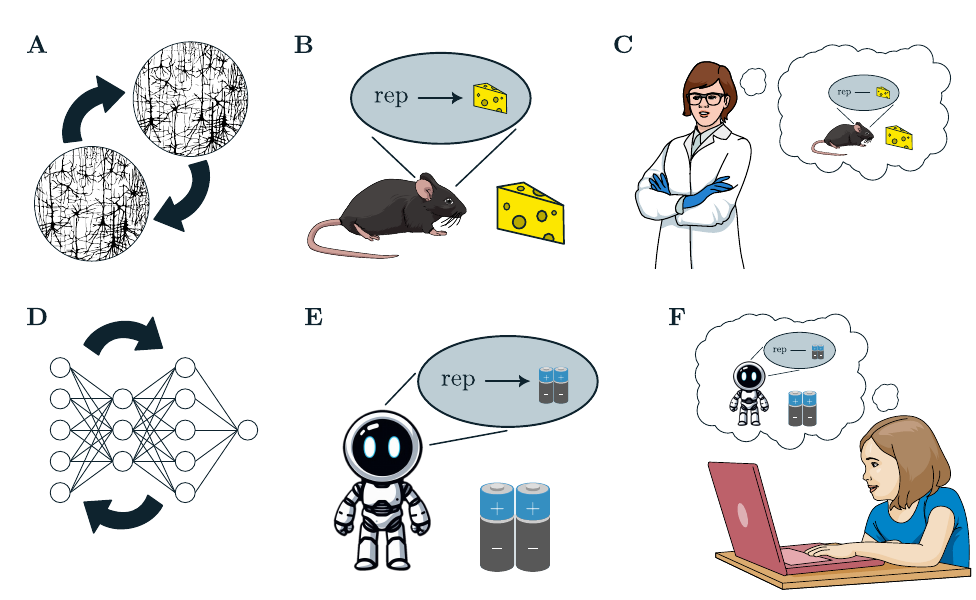}
    \caption[Useful for whom?]{We can think of representations as useful for/usable by another subsystem in the biological or artificial agent \textbf{(A \& D)}, by the agent itself \textbf{(B \& E)}, or by scientists and engineers modeling or designing a system \textbf{(C \& F)}. We primarily discuss A, B, D, \& E, and observe that C and F can (but do not always) reduce to B \& E.}\label{fig:who-is-the-user}
\end{figure}

Conceptualizing representations as being useful and potentially used implies that there is a user of the representation relative to which these terms are applied. This might be worrying: who is this user supposed to be? And does the idea of a user build in assumptions about agency and intentionality, or some kind of homunculus? These questions have sometimes been framed in philosophy in terms of how to identify the ``consumer'' of a representation \citep{millikanLanguageThoughtOther1987, sheaConsumersNeedInformation2007, caoTeleosemanticApproachInformation2012}. In this section we will discuss three types of users of representations that appear in the literature, keeping in mind that just as our goal was not to identify a universal notion of representation, we will not identify a universal notion of the user. And, just as different notions of usefulness, usability, and use can lead to drastically different notions of representation, so can different notions of the user.

In Level 2 above, we began the discussion of usefulness with a discussion of \textit{goals}: representations are useful as part of processes that perform tasks or achieve goals. Within this framework, the user can be understood as the entity or process that performs the task, or has the goals. So a good heuristic for identifying the user of a representation is to start with a particular goal or task: the agent’s goal (surviving, navigating, earning reward), a subsystem’s goal (computing face identity), or even the goal of the scientist, philosopher, or engineer interacting with the agent (explaining or designing it). This will allow us to identify the user as the process or entity that \textit{takes up} the representation in service of the task or goal. We give a few concrete examples below.

\paragraph{The entire system as a user.}

Some conceptions of representations appeal to their usefulness for, or use by, a whole agent in service of some task. For example, we think of a mouse as using internal representations of the spatial layout of its environment for navigating to a desired location (Figure \ref{fig:who-is-the-user}B). This simply means we are understanding the internal states with reference to the mouse-as-a-system’s tasks. That might be because we are interested in their contribution to those tasks but not necessarily the finer-grained processes by which it contributes to them \citep[e.g.,][]{andersonContentActionGuidance2008,rajaEcologicalResonanceReflected2025,chemeroRadicalEmbodiedCognitive2009,dewitAffordancesNeuroscienceSteps2017}. (Some of these authors resist appeals to neural representation, but typically not to weak conceptions of representation, as in Level 1.) So a state may be seen as having useful information, being in a usable format, or being actually used by the whole agent because of its contribution to the agent as a whole accomplishing some task (Figure \ref{fig:who-is-the-user}B,E).

\paragraph{Internal users.}

Some conceptions of representations appeal to their use in a circumscribed computation or subgoal internal to the system. For example, if a system’s task is to navigate safely through a cluttered scene, subgoals may include identifying the type and location of various objects in the scene or planning a route. Neural activity often contributes to the performance of a task by passing through subsequent brain areas or neural network layers, each responding to, transforming, or disentangling the activity of the previous area. In that case we would understand each area or subsystem as the user of the previous area’s representations: it is \textit{their} tasks (sub-tasks of object recognition) with respect to which the representation is understood (Figure \ref{fig:who-is-the-user}A,D).

\paragraph{External users.}

One more possibility is that a representation’s user is a third party outside the agent. Consider two examples. First, it is possible to think of a scientist as the user of a system’s representations. There are tasks — like modeling or explaining the system — that the scientist aims to achieve, and the representations ``feed into'' those tasks by artificial means, e.g., being measured with brain imaging techniques. Second, it may be possible to think of organisms more generally as the users of each other's representations. A bird feigning injury to lead a predator away from its offspring is achieving a task, and doing so using one of the predator’s representations: its representation of the bird as injured prey. If the predator’s representations feed into a system that achieves the bird’s tasks, the bird can be considered a user of its predator’s representations.

In both examples, the representations serving the external agent's tasks do so by serving their more typical tasks. The bird can use the predator’s representations for its task only because the predator is using those representations to identify and pursue prey. Similarly, neuroscience is commonly thought to aim at explaining the way an organism performs its tasks independently of the scientist, and the scientist will have to understand the organism's representations in terms of the tasks it typically uses them for. Some researchers do argue that a computational and representational description of the brain is a problematic imposition from the scientist's perspective onto the organism's \citep{bretteCodingRelevantMetaphor2019}. Others, namely ``active'' or ``haptic'' scientific realists, suggest that these impositions are essential to scientific practice \citep{changWaterH2OEvidence2014, chirimuutaBrainAbstractedSimplification2024}. Nonetheless, for instance, decoding research often restricts itself to decoders that the brain could also implement, because the aim is to determine the content of a representation insofar as it might be used by the brain itself \citep{kriegeskorteInterpretingEncodingDecoding2019,ivanovaLinearRegressionMapping2022}.

Thus, third-party usage does not \emph{necessarily} call for radical changes to our understanding of the representations involved (Figure \ref{fig:who-is-the-user}C,F). That is, a scientist's or engineer's third-party perspective on representations in the brain or robot is often framed in terms of how those representations may be useful for or used by the system under study. This is not to say that the scientist or engineer will always correctly attribute representational capacities, but that a scientist describing the brain \emph{as if} it has representations takes the stance that those putative representations are useful for/usable by/used by the system itself \citep{caoPuttingRepresentationsUse2022, dennettIntentionalStance2002}. There can be, however, cases where an organism’s states are co-opted by a new user in a way that \textit{does} change how we think about their format and content. First, there are many clear cases of scientific observation that re-purpose internal states as indicators of something about the agent other than whatever that state may be used for within the agent. This is the case when blood-oxygen-level in a brain area is used to represent general neural activity, regardless of what the measured neurons themselves represent and without any assumption about the BOLD signal itself serving a representational function. Second, recorded brain activity can be used for various purposes other than what makes the activity significant within the brain. For instance, EEG patterns that originally represent some cognitive content could be read out and used in an authentication system that merely treats them as a unique signature for the individual, or in an artistic tool that uses them as inputs in some creative process, ignoring their original representational content entirely. Third, consider an infant's internal state that corresponds to a state of discomfort or distress and leads to crying; while for the baby this may simply represent its internal condition, within the baby-parent system, the same state functions as a representation of a need for external intervention. In each case, what changes is not just who uses the representation, but what it represents and how its format is understood --- the content and format can shift based on the external user's goals and interpretive framework.

To summarize this subsection: we understand the user of a representation in what philosophers call a \textit{deflationary} sense, as simply \emph{the process or entity that the representation feeds into, and that performs the tasks in question}. Because there are different tasks, because those tasks are achieved by different processes, and because there are different levels at which we can describe and investigate those tasks and processes, there will be different possible ``users'' for any given representation. All this means is that, as we have stressed throughout, the tasks we understand an organism as performing affect the representations we will attribute to it to explain those tasks.

\subsection{Limitations}

Taken together, Levels 1-3 provide a picture of neural representations as internal states that are responsive to variations in ecologically-relevant signals, that are the result of some processing or reformatting, and that are causally involved in downstream behavior. But these different aspects of representation can come apart and, as we’ve stressed, each fits naturally into its own collection of research programs, helping to answer its own set of questions. This raises a question: might representations contribute to a wider range of questions and research programs than we’ve discussed? And to do so, would they have to look different from what Levels 1--3 describe?

For example, representational theories of consciousness, like Higher Order Thought theory, argue that consciousness is explained by representational states \citep{lycan_representational_2017, Rosenthal1997_TheoryConsciousness}. But of course not all representations at Level 1, 2, or 3 are conscious. For instance, visual representations in blindsight patients are understood to be non-conscious, though they can shape behavior. Thus, representational theories of consciousness must rely on a notion of representation that goes beyond our three levels, or rely on an account of how certain conditions imbue representations with additional properties.

Philosophers have also relied on a concept of representation in the study of mind, belief, knowledge, and understanding \citep{kimPhilosophyMind2018, searleMindsBrainsPrograms1998}. Often, it is claimed that genuine or ``original'' representations are what distinguish minds from other sorts of systems \citep{millikanLanguageThoughtOther1987,neanderMarkMentalDefense2017,sheaRepresentationCognitiveScience2018}. But it is plausible that a system could possess representations at Levels 1--3 without also possessing a mind, or without having genuine understanding, beliefs, etc., of the world it represents. To serve as a mark of the mental, or to define qualities like understanding, representations would presumably have to be something more than what we have described here. Representations may even be understood as cognitive phenomenon \textit{to be explained}, as opposed to internal states that \textit{figure into} explanations of cognitive phenomena. 
A related debate plays out in the context of Natural Language Processing, where many researchers agree that Large Language Models do not genuinely understand the language they process, despite possibly having representations of that language in the sense of Levels 1–3 \citep{mitchellDebateUnderstandingAIs2023, benderDangersStochasticParrots2021, titusDoesChatGPTHave2024}. When we attribute to \textit{a person} the ability to represent the semantics of the word ``unicorn,'' we may be describing a capacity to be explained, as opposed to offering an explanation of the capacity in terms of internal ``representations'' and their use.

Our point is neither to promote nor dispute these representationalist approaches to consciousness and mind, but rather to acknowledge the limits of our framework. We have treated the concept of representation throughout this article primarily in terms of what it contributes to a research program, and we want to point out that, among the many research programs using a concept of representation, some seem to ask for more than Levels 1--3 provide. The framework allows us to distinguish the ways that many, but not all researchers rely on representation. Importantly, it also allows us to elucidate those cases where researchers are using a representation concept from different levels, and thus are not disagreeing about what representation is but rather using a different conceptual tool to accomplish their different research goals. We mean to have offered an account of the representation concepts that are prevalent in neuroscience, machine learning, and related areas of philosophy, but we grant that our framework will not capture the representation concepts that exist in other significant areas of research.

\section{Conclusion}

We want to conclude by reflecting on some of the lessons we’ve taken away from this understanding of representation. Since this will have different import for different fields and research programs, we will draw the lessons field-by-field.

\paragraph{For philosophers.}

Many debates remain about the concept of representation and its role in the sciences studying brains, minds, and intelligence. Our framework offers a clarification and new point of entry into some of the most important debates. For example, consider the debate over \textit{representationalism}: whether science \textit{should} conceive of the mind in representational terms. That is an impossible question to answer without a detailed understanding of what the concept of representation contributes to science, and how. And this is what we have tried to clarify by describing how different notions of representation contribute to scientific practice, and what different research programs use them to do. Given the variety of research goals we have said are served by at least one of the three Levels of representational thinking, it would seem a good bet that the representational approach cannot be entirely replaced by alternatives, but we do not claim to have settled the debate over representationalism as a whole. However, an understanding of the wide range of roles that concepts like representation can play in science is a necessary starting point for anyone who would assess the efficacy of using representations to explain what happens in minds and brains \citep[cf. ][]{richmond_commentary_2023, sheaRepresentationCognitiveScience2018, ramseyRepresentationReconsidered2007, bakerThreeAspectsRepresentation2022}. 

Consider also the debate over \textit{representation realism}: whether representations ``really exist'' in brains or machines, or if they are just a useful descriptive language for scientists and engineers. This debate, centering on the question of whether the brain \textit{really has} representations, typically concerns the nature of this special property. What \textit{is} representation? What \textit{makes something count} as a representation? This may be an important question in a context where \textit{really being a representation} is a substantive property that a structure can have or not have --- like the property of really \textit{being responsible} (in the context of assigning punishment), or the property of really \textit{being in pain} (in the context of animal welfare). But in the context of neuroscientific explanation, we have found room for many different notions of representation --- many different ways of being a representation --- sapping this question (which one \textit{really is} representation) of its force. (Though one can still ask a related question, also naturally understood as a question about realism: whether the brain really has the structure attributed to it by any given theory at Levels 1–3.) There is philosophical work to be done concerning what scientists and engineers can and should try to achieve with representational notions, but those questions may look very different from traditional questions about what \textit{makes something count} as a representation \citep{caoPuttingRepresentationsUse2022,chirimuutaBrainAbstractedSimplification2024, eganDeflatingMentalRepresentation2025, richmondWhatTheoryNeural2025}.

\paragraph{For neuroscientists and cognitive scientists.}

Our breakdown of the different aspects of use and usability, and the three levels, might sound familiar, as it echoes decades of debates and opinions on the nature of the neural code \citep{fodorLanguageThought1975,bechtelRepresentationsCognitiveExplanations1998,parkerSenseSingleNeuron1998,edelmanRepresentationRepresentationSimilarities1998,sheaRepresentationCognitiveScience2018,ritchieDecodingBrainNeural2019}. Still, there are things to be gained from broadening one's conception of neural representations to encompass all three levels and the various aspects of use and usability. This can reveal blind spots in our methodologies and suggest new research. For instance, studies looking at neural tuning properties are widespread and typically take a Level-1 approach. We have also observed that almost all existing quantitative tools for \emph{comparing} representations across systems live at Level 1 or Level 2. This suggests that neuroscience as a field may have a blind spot --- and opportunity for future work --- in terms of expanding tools for studying how representations are used (cf. \cite{bansalRevisitingModelStitching2021}). Our discussion in Level 2 above also invites a broad perspective on what is ``useful'' or ``usable,'' and serves as a reminder that conclusions we draw about representations are often inextricable from the assumptions we make about the goals and capabilities of the system under study, assumptions which are often baked implicitly into the methods we use.

\paragraph{For computer scientists.}

Broadly speaking, the field of computer science enjoys a particularly wide range of concepts and tools for studying ``representations.'' Neural network researchers have white-box access to all aspects of their models and can perform arbitrary interventions, but this does not trivialize the problem of designing or analyzing representations in our ML or AI systems. What makes one representation better than another --- the purview of the sub-field of representation learning --- is still an active research area at Level 1 and Level 2. Meanwhile, researchers aiming to uncover and interpret the inner-workings of large machine learning models (``mechanistic interpretability'') have sought more and more to explain the behavior of those models using Level-3 tools \citep{lindsey2025biology, ameisen2025circuit, templetonScalingMonosemanticityExtracting2024,liEmergentWorldRepresentations2024}. The main take-home message for computer science is to embrace the plurality of concepts of representations and to be precise about which particular conception of ``representation'' (which Level) is being used in a given context. Finally, it is important for computer scientists to keep an open dialogue with neuroscience and philosophy because there are notions of representations that go beyond what we could cover here.

\section*{Acknowledgements}

This paper is the result of a Generative Adversarial Collaboration that began at the Cognitive Computational Neuroscience conference in 2021, originally titled ``What makes representations ‘useful’?''. Special thanks to Nick Shea and Alessandro Achille for contributions to our original workshop and early discussions. Thanks to Timothy Schroeder and Chris Eliasmith for helpful conversations on {\it mis}representations. Thanks also to the CCN GAC organizers for facilitating this interdisciplinary project. We acknowledge support from NSF grant 2229929 (XP, NK).

\bibliography{zotero2}

\end{document}